\documentclass[10pt,aps,onecolumn]{revtex4}
\usepackage{epsfig}
\usepackage{amsmath}
\usepackage{bm}
\usepackage{epsfig}
\usepackage{epstopdf}
\usepackage{subfig}
\usepackage{graphicx}
\usepackage[section]{placeins}
\usepackage[toc]{appendix}
\begin{document}
\title{Enhancement and Suppression of Four wave mixing in coupled semiconductor quantum dot-gold grating systems.}
\author{Shailendra Kumar Singh}
\email{singhshailendra3@gmail.com}
\author{Mehmet Emre Tasgin}
\email{metasgin@hacettepe.edu.tr}
\affiliation{Institute of Nuclear Sciences, Hacettepe University, 06800,Ankara, Turkey}
\begin{abstract}
We have shown that Four wave mixing (FWM) processes of electromagnetic field modes of a grating can be controlled by the presence of interactions with  a quantum dot or a molecule made by coupled quantum dots. By choosing the appropritae level spacing for the quantum emitter, one can either suppress or enhance the Four wave mixing process. We revel theoretically the underlying mechanism for this effect.
(i) Suppression in FWM intensity occurs simply because induced Electromagnetic Induced Transparency does not allow the excitation at converted FWM frequency.
(ii) Enhancement emerges since FWM process can be brought to resonance. Path interference effect cancels the nonresonant frequency terms. Furthermore, we have also shown that in case of coupled quantum dots enhancement increases significantly as compared to the case of a single quantum dot.

\end{abstract}
\maketitle
\section{Introduction}
Quantum Plasmonics is an emerging area of research which involves the study of the optical properties of hybrid photonic structures incorporating both plasmonic nanostructures and quantum emitters \cite{Tame}, such as atoms, molecules and semiconductor quantum dots. These complex hybrid  are active photonic structures and expected to enhance optical response significantly, for example modification of the linear susceptibility \cite{Lu,Hatef,Sadeghi,Paspalakis, Tasgin} and the enhancement of nonlinear susceptibilities in several quantum systems with different level structures coupled to various plasmonic nanostrcutures \cite{Lu1,Pu,Paspalakis1,Singh,Paspalakis2}.\

Four Wave Mixing is one of the above mentioned  nonlinear process of light-matter interactions in which three incoming waves,indicated as $\omega_{1}$, $\omega_{2}$, $\omega_{3}$ in the material generate a fourth wave of frequency $\omega_{4}$ \cite{Yurke}. Assuming that the three incident waves have frequencies in the visible or near-infrared range, the incoming electric fields $E\left(\omega_{i}\right)$ $($with $i=1,2,3)$ interact with the material's electrons to induce a nonlinear polarization $P^{\left(3\right)}\left( \omega_{4}\right)$ in the illuminated volume. The magnitude of the polarization is determined by the strength of the incident fields and the efficiency with which the material can be
polarized \cite{Wang}. The latter is indicated with the third-order nonlinear susceptibility
$\chi^{\left(3\right)}$, a measure of the material's response to the incoming fields.
Four wave mixing (FWM) has found numerous practical applications,
including: optical processing; nonlinear imaging; real-time holography
and phase-conjugate optics; phase-sensitive amplification; and entangled photon
pair production \cite{Scully}.\

In several recent studies, the modification of $\chi^{\left(3\right)}$ (FWM process) susceptibility in a quantum dot system coupled to spherical nanoparticle has been investigated when the hybrid structures interacts with a weak probe field and a strong pump field \cite{Lu1,Li,Liu,Li1}. All these works have shown for different distance between the quantum dot and the metal nanoparticle the $\chi^{\left(3\right)}$ susceptibility can be either enhanced or strongly suppressed. In addition, bistable behavior has been also reported in these kind of systems \cite{Paspalakis2,Li}.\\

Here, we propose a method for increasing the efficiency of FWM processes by exploiting gold
grating \cite{Jan,Poutrina}. Narrow peaks are observed in the transmission spectra of p-polarized light passing through a thin gold film that is coated on the surface of a transparent diffraction grating. The spectral position and intensity of these peaks can be tuned over a wide range of wavelengths by simple rotation of the grating \cite{Bipin}. The wavelengths where these transmission peaks are observed correspond to conditions where surface plasmon resonance occurs at the gold-air interface. Light diffracted by the grating couples with surface plasmons in the metal film to satisfy the resonant condition, resulting in enhanced light transmission through the film.\

The paper is organized as follows. In Section II, we describe  the FWM Process  in the coupled system of gold grating with a quantum oscillator. In the same section, we introduce the Hamiltonian for hybrid system. FWM process is also included in the second quantized Hamiltonian. We derive the equations of motion for the system using the density matrix formalism for the quantized quantum oscillator. We use phenomenological way to include damping of gold grating modes as well as quantum emitter. In Section II B, We demonstrated that FWM process can be suppressed for  $\left(
\omega _{eg}=2\omega -\omega^{\prime}\right)$. In Section II C, we present a contrary effect where FMW process can be enhanced. This is due to cancellation of non resonant terms in denominator. In Section III, we further investigate the case of gold grating coupled to two quantum emitters (both quantum emitters coupled to each other also) simultaneously. We conclude our results in Section IV.
 
\section{Hamiltonian for Four Wave Mixing}

The total Hamiltonian $\hat{H}$~\ for the described system can be written as
Sum of the energy of the Quantum Oscillator $\hat{H}_{0}$,( In our case we
have taken a QD of energy levels $\left\vert e\right\rangle ~$and $%
\left\vert g\right\rangle $) , enegy of the elctromagnetic modes of Gold
Grating $\left( \hat{a}_{1},\hat{a}_{2},\hat{a}_{3}\right) $ for a
particular angle of incidence of Pump lasers $\hat{H}_{grating}$, the
interaction of the Quantum Oscillator with the Grating Modes $\hat{H}_{int}$

\begin{equation}
\hat{H}_{0}=\hbar \omega _{e}\left\vert e\right\rangle \left\langle
e\right\vert +\hbar \omega _{g}\left\vert g\right\rangle \left\langle
g\right\vert
\end{equation}
\begin{equation}
\hat{H}_{grating}=\hbar \omega _{1}\hat{a}_{1}^{\dagger }\hat{a}_{1}+\hbar
\omega _{2}\hat{a}_{2}^{\dagger }\hat{a}_{2}+\hbar \omega _{3}\hat{a}%
_{3}^{\dagger }\hat{a}_{3}
\end{equation}
\begin{equation}
\hat{H}_{int}=\hbar \left( f~\hat{a}_{3}^{\dagger }\left\vert g\right\rangle
\left\langle e\right\vert +f~^{\ast }\hat{a}_{3}\left\vert e\right\rangle
\left\langle g\right\vert \right)
\end{equation}
Here, we have considered that level spacing of the QD is only resonant to $%
\hat{a}_{3}$ mode (i.e. $\omega _{eg}\sim \omega _{3}$). as well as the
energy transferred by the pump source $\omega $ and $\omega ^{\prime }.$%
\begin{equation}
\hat{H}_{P}=i\hbar \left( \hat{a}_{1}^{\dagger }~\epsilon _{P}e^{-i\omega t}-%
\hat{a}_{1}~\epsilon _{P}^{\ast }e^{i\omega t}\right) +i\hbar \left( \hat{a}%
_{2}^{\dagger }~\epsilon _{P}^{\prime }e^{-i\omega ^{\prime }t}-\hat{a}%
_{2}~\epsilon _{P}^{\ast \prime }e^{i\omega ^{\prime }t}\right)
\end{equation}
\begin{equation}
\hat{H}_{FWM}=\hbar \chi ^{(2)}\left( \hat{a}_{3}^{\dagger }\hat{a}%
_{2}^{\dagger }\hat{a}_{1}^{2}+\hat{a}_{1}^{\dagger 2}\hat{a}_{2}\hat{a}%
_{3}~\right)
\end{equation}
For the Process of $\left( \omega _{3}=~2\omega _{1}-\omega _{2}\right) ~$as
mentioned in PRL 103, 266802.

In Eq. (1), $\hbar \omega _{e}$ ($\hbar \omega _{g}$) is the excited
(ground) state energy of the Quantum Oscillator. \ States $\left\vert
e\right\rangle $ and $\left\vert g\right\rangle $ corresponds to excited and
ground levels of the Quantum Oscillator respectively. \ $\left( \hat{a}_{1},%
\hat{a}_{2},\hat{a}_{3}\right) $ are the Gold Grating modes at a particulat
angle of incidence $\theta =5^{\circ }$. $f~$\ is the coupling matrix
element between the field of grating mode and the Quantum Oscillator. Eq.
(4) describes the driving the Electromagnetic field modes of grating $\left( 
\hat{a}_{1}\text{ and }\hat{a}_{2}\right) $ with $e^{-i\omega t}$ and $%
e^{-i\omega ^{\prime }t}$ respectively.

Eq.(5) describes where the Four wave mixing takes place in which $\hat{a}%
_{1} $ mode contributes two photons and $\hat{a}_{2}$ mode single photons in
the process.

\subsection{ Heisenberg Equations of Motion}

We use the commutation relations 
\begin{equation}
i\hbar \frac{d}{dt}\hat{O}=\left[ \hat{O},\hat{H}\right]
\end{equation}
for deriving equations of motions. After obtaining the dynamics in the
quantum approach, we carry $\left( \hat{a}_{1},\hat{a}_{2},\hat{a}%
_{3}\right) ~$to classical expectation values $\left( \alpha _{1},\alpha
_{2},\alpha _{3}\right) .~$We also introduce the decay rates for $\left(
\alpha _{1},\alpha _{2},\alpha _{3}\right) .$ Quantum Oscillator is treated
within the density matrix approach. The equations of motion take the form 
\begin{subequations}
\begin{equation}
\dot{\alpha}_{1}=\left( -i\omega _{1}-\gamma _{1}\right) \alpha _{1}-2i\chi
^{\left( 3\right) }\alpha _{1}^{\ast }\alpha _{2}\alpha _{3}+\epsilon
_{p}e^{-i\omega t}  \label{7a}
\end{equation}
\begin{equation}
\dot{\alpha}_{2}=\left( -i\omega _{2}-\gamma _{2}\right) \alpha _{2}-i\chi
^{\left( 3\right) }\alpha _{3}^{\ast }\alpha _{1}^{2}+\epsilon _{p}^{\prime
}e^{-i\omega ^{\prime }t}  \label{7b}
\end{equation}
\begin{equation}
\dot{\alpha}_{3}=\left( -i\omega _{3}-\gamma _{3}\right) \alpha _{3}-i\chi
^{\left( 3\right) }\alpha _{2}^{\ast }\alpha _{1}^{2}-if\rho _{ge}
\label{7c}
\end{equation}
\begin{equation}
\dot{\rho}_{ge}=\left( -i\omega _{eg}-\gamma _{eg}\right) \rho
_{ge}+if\alpha _{3}\left( \rho _{ee}-\rho _{gg}\right)  \label{7d}
\end{equation}
\begin{equation}
\dot{\rho}_{ee}=-\gamma _{ee}\rho _{ee}+~if\left( \alpha _{3}^{\ast }\rho
_{ge}-\alpha _{3}\rho _{eg}\right)  \label{7e}
\end{equation}
where $\gamma _{1}$,$\gamma _{2},\gamma _{3}~$are the damping rates of the
electromagnetic modes of the gold grating $\left( \alpha _{1},\alpha
_{2},\alpha _{3}\right) .~\gamma _{ee}~$and $\gamma _{eg}=$ $\gamma _{ee}/2~$%
are the diagonal and off-diagonal elements of the quantum oscillator
respectively. The constraints of the conservation probability $\rho
_{ee}+~\rho _{gg}=1$ accompanies above set of equations.

Besides the time-evolution simulations, one may gain the understanding by
seeking solutions of the following form. For long time behavior we take
solutions of the form

$\alpha _{1}(t)=\tilde{\alpha}_{1}e^{-i\omega t},\alpha _{2}(t)=\tilde{\alpha%
}_{2}e^{-i\omega ^{\prime }t},\alpha _{3}(t)=\tilde{\alpha}_{3}e^{-i\left(
2\omega -\omega ^{\prime }\right) t}$ (Condition of Four wave mixing
process), $\rho _{ge}=\tilde{\rho}_{ge}e^{-i\left( 2\omega -\omega ^{\prime
}\right) t}$ here we have considered that level spacing of the QD is only
resonant to $\hat{a}_{3}$ mode (i.e. $\omega _{eg}\sim \omega _{3}$), $\rho
_{ee}=\tilde{\rho}_{ee}.$

Inserting the solutions in above set of equations (7a-7e) we have for long
time behavior 
\end{subequations}
\begin{subequations}
\begin{equation}
\left[ i\left( \omega _{1}-\omega \right) +\gamma _{1}\right] \tilde{\alpha}%
_{1}+2i\chi ^{\left( 3\right) }\tilde{\alpha}_{1}^{\ast }\tilde{\alpha}_{2}%
\tilde{\alpha}_{3}=\epsilon _{p}  \label{8a}
\end{equation}
\begin{equation}
\left[ i\left( \omega _{2}-\omega ^{\prime }\right) +\gamma _{2}\right] 
\tilde{\alpha}_{2}+i\chi ^{\left( 3\right) }\tilde{\alpha}_{3}^{\ast }\tilde{%
\alpha}_{1}^{2}=\epsilon _{p}^{\prime}  \label{8b}
\end{equation}
\begin{equation}
\left[ i\left( \omega _{3}+\omega ^{\prime }-2\omega \right) +\gamma _{3}%
\right] \tilde{\alpha}_{3}+i\chi ^{\left( 3\right) }\tilde{\alpha}_{2}^{\ast
}\tilde{\alpha}_{1}^{2}=-if\tilde{\rho}_{ge}  \label{8c}
\end{equation}
\begin{equation}
\left[ i\left( \omega _{eg}+\omega ^{\prime }-2\omega \right) +\gamma _{eg}%
\right] \tilde{\rho}_{ge}=if\tilde{\alpha}_{3}\left( \tilde{\rho}_{ee}-%
\tilde{\rho}_{gg}\right)  \label{8d}
\end{equation}
\begin{equation}
\gamma _{ee}\tilde{\rho}_{ee}=if\left( \tilde{\alpha}_{3}^{\ast }\tilde{\rho}%
_{ge}-\tilde{\alpha}_{3}\tilde{\rho}_{eg}\right)  \label{8e}
\end{equation}
Using equations (8c) and (8d), we obtain the steady state value for $\tilde{%
\alpha}_{3}$ as follows 
\end{subequations}
\begin{equation}
\tilde{\alpha}_{3}=\frac{i\chi ^{\left( 3\right) }\tilde{\alpha}_{2}^{\ast }%
\tilde{\alpha}_{1}^{2}}{\frac{\left\vert f\right\vert ^{2}y}{\left[ i\left(
\omega _{eg}+\omega ^{\prime }-2\omega \right) +\gamma _{eg}\right] }-\left[
i\left( \omega _{3}+\omega ^{\prime }-2\omega \right) +\gamma _{3}\right] }
\end{equation}
Where $y=\left( \tilde{\rho}_{ee}-\tilde{\rho}_{gg}\right)$is the steady
state value of the population inversion. If the quantum oscillator is tuned
around $\omega _{eg}=2\omega -\omega ^{\prime },$ $\tilde{\alpha}_{3}$ can
be suppressed.

\begin{figure}[hbt!]
\centering
{\includegraphics[width=0.5\textwidth]{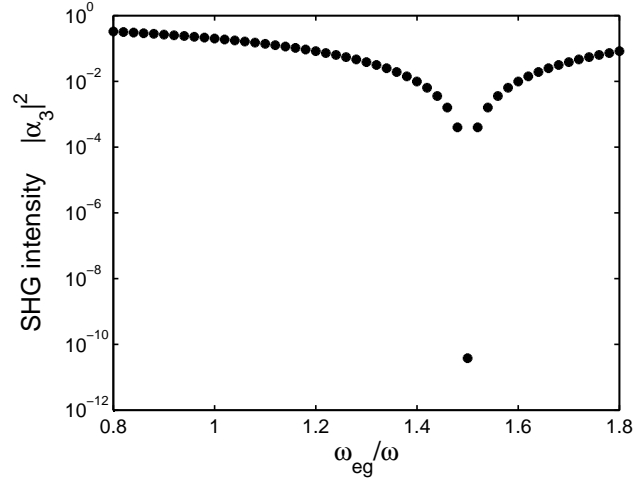}}
\caption{Suppression of the FWM intensity to the $\hat{a}_{3}$ gold grating
mode from the $\hat{a}_{2}$ and $\hat{a}_{1}$ mode. Even at the presence of
the resonant FWM condition, $\protect\omega _{1}= \protect\omega=1.0$, $%
\protect\omega _{2}= \protect\omega^{\prime}=0.5$ and $\left(\omega
_{3}=2\omega -\omega^{\prime}\right)$, the presence of
quantum oscillator prevents to take place of the FWM process. EIT does not
allow the FWM process. The resonant FWM conversion is represented by unity
in figure. When $\left(\protect\omega _{eg}=2\protect\omega -\protect\omega%
^{\prime}\right)$, the FWM intensity even can be suppressed by 10 orders of
magnitude with respect to resonant value. Decay rates for our numerical
simulations are $\protect\gamma _{1}=\protect\gamma _{2}=\protect\gamma_{3}=
0.01\protect\omega$ and $\protect\gamma _{eg}= 0.00001\protect\omega$. We
have taken $\protect\chi^{\left(2\right)} = 0.00001\protect\omega$ and $f
=0.1\protect\omega$}
\label{Fig 1}
\end{figure}

\begin{figure}[hbt!]
\centering
{\includegraphics[width=0.5\textwidth]{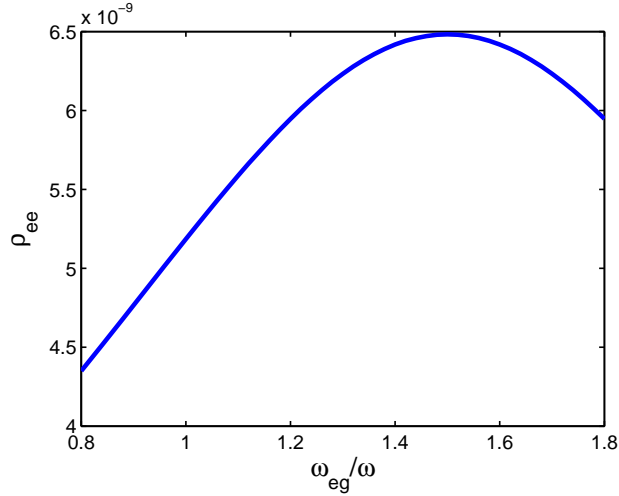}}
\caption{Enhancement of population in excited level of the quantum
oscillator coupled to gold grating at the resonance condition $\left(\protect%
\omega _{eg}=2\protect\omega -\protect\omega^{\prime}\right)$. We can see
the population in excited level is maximum at this condition unlike FWM
Intensity $\vert$ $\tilde{\protect\alpha}_{3}$ $\vert^{2}$ shown in Fig. 1.
All other parameters for numerical simulation remain same like in Fig 1.}
\label{Fig 2}
\end{figure}

\subsection{Suppression of the Four wave Mixing Process}

We can see from Eq. (9) that $\left\vert f\right\vert ^{2}y/\gamma _{eg}$
can attain huge values on resonance $\omega _{eg}=~2\omega -\omega ^{\prime }
$as well as linewidth of the quantum oscillator $\gamma _{eg}$ is very small
\ compared to the all other frequencies. If $f\neq 0$, the largeness of the $%
\left\vert f\right\vert ^{2}y/\gamma _{eg}$ term dominates the denominator.
This results in the suppression of the generation of the FWM mode $\tilde{%
\alpha}_{3}$ in our model Hamiltonian system. In Fig.1 we have shown that
FWM process can be suppressed very effectively by coupling to gold grating
to quantum oscillator. We have time evolve Eqs. $\left(7a-7e\right)$ to
obtain steady state values for the FWM intensity.\newline
Without the presence of quantum oscillator, the FWM would be maximum $\tilde{
\alpha}_{3}=-\frac{i\chi ^{\left( 3\right) }\tilde{\alpha}_{2}^{\ast} \tilde{%
\alpha}_{1}^{2}}{\gamma _{3}}$ when the FWM mode is on resonance $\left(
\omega _{3}=2\omega -\omega ^{\prime}\right)$. In Fig. 1 we observe that
even at the presence of this resonance condition $\left(\omega _{3}=2\omega
-\omega^{\prime}\right)$, EIT suppresses the FWM by 10 order of magnitude.
Furthermore, in this case population of excited level of quantum oscillator
is maximum at this point as shown in Fig.2 as well as population inversion
is approximately $y=\left( \tilde{\rho}_{ee}-\tilde{\rho}_{gg}\right)\approx
-1$

\subsection{Enhancement of Four Wave Mixing Process}

Similar to suppression phenomena, the interference effects can be arranged
in such a way that FWM process can be carried closer to resonance. In the
denominator of Eq.(9), \ the imaginary part of the first term$~\frac{%
\left\vert f\right\vert ^{2}y}{\left[ i\left( \omega _{eg}+\omega ^{\prime
}-2\omega \right) +\gamma _{eg}\right] }$ can be arranged to cancel the$%
~i\left( \omega _{3}+\omega ^{\prime }-2\omega \right) $ factor in the
second term of the denominator. This gives the condition 
\begin{equation}
\left\vert f\right\vert ^{2}y\left( \omega _{eg}+\omega ^{\prime }-2\omega
\right) +\left( \omega _{3}+\omega ^{\prime }-2\omega \right) \left[ \left(
\omega _{eg}+\omega ^{\prime }-2\omega \right) ^{2}+\gamma _{eg}^{2}\right]
=0
\end{equation}
Eq.(10) has two roots. 
\begin{equation}
\left( \omega _{eg}^{\left( 1,2\right) }+\omega ^{\prime }-2\omega \right) =%
\frac{\left\vert f\right\vert ^{2}y}{\left( \omega _{3}+\omega ^{\prime
}-2\omega \right) }\mp \sqrt{\frac{\left\vert f\right\vert ^{4}y^{2}}{\left(
\omega _{3}+\omega ^{\prime }-2\omega \right) ^{2}}-4\gamma _{eg}^{2}}
\end{equation}
The first smaller root $\omega _{eg}^{\left( 1\right)}\approx 2\omega
-\omega ^{\prime}$ is not very useful for FWM enhancement, as it enhance the
real part of the term $~\frac{\left\vert f\right\vert ^{2}y}{\left[
i\left(\omega _{eg}+\omega ^{\prime }-2\omega \right) +\gamma _{eg}\right]} ~
$ to rapidly diverge as we have seen in suppression condition for FWM,
whereas $\omega _{eg}^{\left(2\right)}$ minimizes the absolute value of the
denominator of Eq. (9) that gives enhancement of FWM process. For the case
of suppression of FWM, one can safely use the approximation $y$ $\approx -1$
because excitations are suppressed in the hybrid system $\rho _{ee}\approx 0$
and this leads to $y$ =$\left(\rho _{ee}-\rho _{gg}\right) \approx -1$.
However, in case of FWM enhancement, one can not approximate $y$ $\approx -1$%
. Nevertheless, Eq.(11) still serves at least a guess value for the order of 
$\omega _{eg}^{\left( 2\right)}$, where FWM enhancement arises. 
\begin{figure}[hbt!]
\centering
{\includegraphics[width=0.5\textwidth]{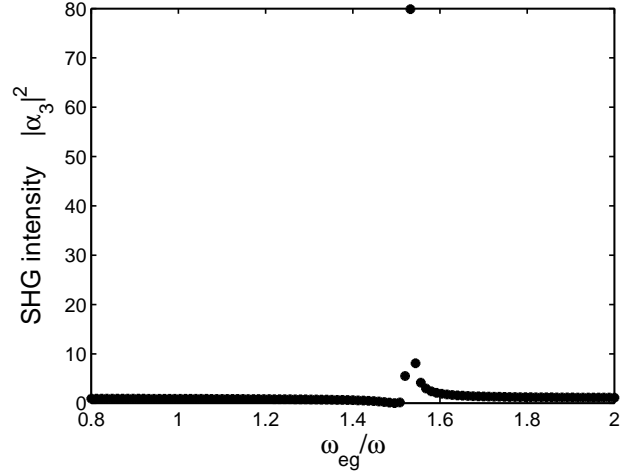}}
\caption{The enhancement of the FWM process. The FWM mode $\hat{a}_{3}$ is
far-off resonant to the FWM condition $\left(\protect\omega _{3}= 1.85\omega\right)$. The FWM process can be carried closer to resonance by arranging the quantum level spacing to $\left(\omega _{eg}\approx 1.52\omega\right)$. The conversion is enhanced nearly $80$ times compared to off-resonant process. The conversion for off resonant process $\left( f = 0\right)$ is represented by unity in figure. For $\left( 
\omega _{eg}=2\omega -\omega^{\prime}\right)$, FWM
process is suppressed similar to Fig.1. Decay rates of the grating modes are
taken as $\gamma _{1}=\gamma _{2}=\gamma_{3}=0.01\omega$. We use $\chi^{\left(2\right)} = 0.00001
\omega$ and $f=0.1\omega$ for numerical simulations.}
\label{Fig 3}
\end{figure}

\section{Two Coupled Quantum Dots}

In case of two coupled QDs we have total Hamiltonian as Follows. 
\begin{equation}
\hat{H}_{0}=\hbar \omega _{eg}^{(1)}\left\vert e_{1}\right\rangle
\left\langle e_{1}\right\vert +\hbar \omega _{eg}^{(2)}\left\vert
e_{2}\right\rangle \left\langle e_{2}\right\vert
\end{equation}
\begin{equation}
\hat{H}_{grating}=\hbar \omega _{1}\hat{a}_{1}^{\dagger }\hat{a}_{1}+\hbar
\omega _{2}\hat{a}_{2}^{\dagger }\hat{a}_{2}+\hbar \omega _{3}\hat{a}%
_{3}^{\dagger }\hat{a}_{3}
\end{equation}
\begin{equation}
\hat{H}_{int}=\hbar \left[ \left(f_{1}\hat{a}_{3}^{\dagger }\left\vert
g_{1}\right\rangle \left\langle e_{1}\right\vert +f_{1}^{\ast }\hat{a}%
_{3}\left\vert e_{1}\right\rangle \left\langle g_{1}\right\vert \right)
+\left( f_{2}\hat{a}_{3}^{\dagger }\left\vert g_{2}\right\rangle
\left\langle e_{2}\right\vert +f_{2}^{\ast }\hat{a}_{3}\left\vert
e_{2}\right\rangle \left\langle g_{2}\right\vert \right) \right]
\end{equation}
Here, we have considered that level spacing of both the QDs is only resonant
to $\hat{a}_{3}$ mode (i.e. $\omega _{eg}^{(1)}\sim \omega _{3}$, $\omega
_{eg}^{(2)}\sim \omega _{3}$). 
\begin{equation}
\hat{H}_{QE-QE}=\hbar \left[ g\left( \left\vert e_{2}\right\rangle
\left\langle g_{2}\right\vert \otimes \left\vert g_{1}\right\rangle
\left\langle e_{1}\right\vert \right) +g^{\ast }\left( \left\vert
e_{1}\right\rangle \left\langle g_{1}\right\vert \otimes \left\vert
g_{2}\right\rangle \left\langle e_{2}\right\vert \right) \right]
\end{equation}
as well as energy transferred by Pump Source with $e^{-i\omega t}$ and $%
e^{-i\omega ^{\prime}t}$ respectively. 
\begin{equation}
\hat{H}_{P}=i\hbar \left( \hat{a}_{1}^{\dagger }~\epsilon _{P}e^{-i\omega t}-%
\hat{a}_{1}~\epsilon _{P}^{\ast }e^{i\omega t}\right) +i\hbar \left( \hat{a}%
_{2}^{\dagger }~\epsilon _{P}^{\prime }e^{-i\omega ^{\prime }t}-\hat{a}%
_{2}~\epsilon _{P}^{\ast \prime }e^{i\omega \prime t}\right)
\end{equation}
\begin{equation}
\hat{H}_{FWM}=\hbar \chi ^{(3)}\left( \hat{a}_{3}^{\dagger }\hat{a}%
_{2}^{\dagger }\hat{a}_{1}^{2}+\hat{a}_{1}^{\dagger 2}\hat{a}_{2}\hat{a}%
_{3}~\right)
\end{equation}
is the Four Wave Mixing Hamiltonian for the Process of $\left(\omega
_{3}=~2\omega _{1}-\omega _{2}\right)$ as mentioned in PRL 103, 266802.

By using the commutation relation Eq.(6) as well as proceeding like the same
way for a single QD case, we get following equations for the case of coupled
QDs. 
\begin{subequations}
\begin{equation}
\dot{\alpha}_{1}=\left( -i\omega _{1}-\gamma _{1}\right) \alpha _{1}-2i\chi
^{\left( 3\right) }\alpha _{1}^{\ast }\alpha _{2}\alpha _{3}+\epsilon
_{p}e^{-i\omega t}  \label{18a}
\end{equation}
\begin{equation}
\dot{\alpha}_{2}=\left( -i\omega _{2}-\gamma _{2}\right) \alpha _{2}-i\chi
^{\left( 3\right) }\alpha _{3}^{\ast }\alpha _{1}^{2}+\epsilon _{p}^{\prime
}e^{-i\omega ^{\prime }t}  \label{18b}
\end{equation}
\begin{equation}
\dot{\alpha}_{3}=\left( -i\omega _{3}-\gamma _{3}\right) \alpha _{3}-i\chi
^{\left( 3\right) }\alpha _{2}^{\ast }\alpha _{1}^{2}-if_{1}\rho
_{ge}^{\left( 1\right) }-if_{2}\rho _{ge}^{\left( 2\right)}  \label{18c}
\end{equation}
\begin{equation}
\dot{\rho}_{ge}^{\left( 1\right) }=\left(-i\omega _{eg}^{\left(1\right)
}-\gamma _{eg}^{\left(1\right)}\right) \rho _{ge}^{\left( 1\right)
}+if_{1}^{\ast }\alpha _{3}\left( \rho _{ee}^{\left( 1\right) }-\rho
_{gg}^{\left( 1\right) }\right) +ig^{\ast }\left( \rho _{ee}^{\left(
1\right) }-\rho _{gg}^{\left( 1\right) }\right) \rho _{ge}^{\left( 2\right)}
\label{18d}
\end{equation}
\begin{equation}
\dot{\rho}_{ge}^{\left(2\right)}=\left(-i\omega _{eg}^{\left(2\right)
}-\gamma _{eg}^{\left(2\right)}\right) \rho _{ge}^{\left(2\right)
}+if_{2}^{\ast}\alpha _{3}\left(\rho _{ee}^{\left( 2\right)}-\rho
_{gg}^{\left(2\right)}\right) +ig\left( \rho _{ee}^{\left(2\right)}-\rho
_{gg}^{\left(2\right)}\right) \rho _{ge}^{\left(1\right)}  \label{18e}
\end{equation}
\begin{equation}
\dot{\rho}_{ee}^{\left( 1\right) }= -\gamma _{ee}^{\left( 1\right)}\rho
_{ee}^{\left( 1\right)}+i\left(f_{1}\alpha _{3}^{\ast}\rho _{ge}^{\left(
1\right) }-f_{1}^{\ast}\alpha _{3}\rho _{eg}^{\left( 1\right)}\right)
+i\left( g\rho _{eg}^{\left(2\right) }\rho _{ge}^{\left( 1\right)}-g^{\ast
}\rho _{eg}^{\left( 1\right)}\rho _{ge}^{\left( 2\right)}\right)  \label{18f}
\end{equation}
\begin{equation}
\dot{\rho}_{ee}^{\left( 2\right)}= -\gamma _{ee}^{\left(2\right)}\rho
_{ee}^{\left(2\right)} + i\left(f_{2}\alpha _{3}^{\ast }\rho _{ge}^{\left(
2\right) }-f_{2}^{\ast }\alpha _{3}\rho _{eg}^{\left(2\right)}\right)
+i\left( g^{\ast }\rho _{eg}^{\left( 1\right) }\rho _{ge}^{\left(2\right)
}-g\rho _{eg}^{\left(2\right)}\rho _{ge}^{\left(1\right)}\right)  \label{18g}
\end{equation}
\end{subequations}
where $\gamma _{1}$,$\gamma _{2},\gamma _{3}~$are the damping rates of the
electromagnetic modes of the gold grating $\left( \alpha _{1},\alpha
_{2},\alpha _{3}\right) .~~\gamma _{ee}^{\left( 1\right) }$, $\gamma
_{ee}^{\left( 2\right) }$ and $~\gamma _{eg}^{\left( 1\right) }=\gamma
_{ee}^{\left( 1\right) }/2,~\gamma _{eg}^{\left( 2\right) }=\gamma
_{ee}^{\left( 2\right) }/2,~$are the diagonal and off-diagonal decay rates
of the first and second quantum emitter respectively. The constraints of the
conservation probability $\rho _{ee}^{\left( 1\right) }+~\rho _{gg}^{\left(
1\right) }=1$ and $\rho _{ee}^{\left( 2\right) }+~\rho _{gg}^{\left(
2\right) }=1$ accompanies above set of Eqs.(18a-18g).

In our simulation for enhancement process of FWM, we time evolve Eqs.
(18a-18g) numerically to obtain the long time behaviors of $%
\rho_{ge}^{\left( 1\right)},\rho _{ge}^{\left( 2\right)},\rho
_{ee}^{\left(1\right)},\rho _{ee}^{\left( 2\right)},\alpha _{1},\alpha _{2}~$
and $\alpha _{3}$. We determine the values to where they converge when the
drive is on for long enough times. We perform this simulations for different
parameter sets $\left( f_{1},f_{2},g,\omega _{eg}^{\left( 1\right)},\omega
_{eg}^{\left( 2\right)},\gamma _{eg}^{\left( 1\right)},\gamma _{eg}^{\left(
2\right)}\right)$ with the initial condition $\rho _{ee}^{\left(
1\right)}\left( t=0\right) =\rho _{ee}^{\left(2\right) }\left( t=0\right)
=0, $ $\rho _{ge}^{\left( 1\right)}\left(t=0\right) =\rho
_{ge}^{\left(2\right)}\left(t=0\right) =0,~\alpha _{1}\left(
0\right)=0,~\alpha _{2}\left( 0\right) =0,~\alpha _{3}\left( 0\right) =0$.%
\newline

Besides the time-evolution simulations, one may gain the understanding by
seeking the solutions of the following form:

$\alpha _{1}(t)=\tilde{\alpha}_{1}e^{-i\omega t}$, $\alpha _{2}(t)=\tilde{%
\alpha }_{2}e^{-i\omega ^{\prime }t}$, $\alpha _{3}(t)=\tilde{\alpha}%
_{3}e^{-i\left( 2\omega -\omega ^{\prime }\right)t}$ (Condition of Four wave
mixing process), $\rho _{ge}^{\left( 1\right)}=\tilde{\rho}_{ge}^{\left(
1\right) }e^{-i\left( 2\omega -\omega ^{\prime }\right)t}$, $\rho
_{ge}^{\left( 2\right)}=\tilde{\rho}_{ge}^{\left( 2\right) }e^{-i\left(
2\omega -\omega ^{\prime}\right)t}$.

Here we have considered that level spacing of both the QDs is only resonant
to $\hat{a}_{3}$ mode (i.e. $\omega _{eg}^{\left( 1\right)}\sim \omega
_{3},\omega _{eg}^{\left(2\right)}\sim \omega _{3})$, $\rho _{ee}^{\left(
1\right)}(t)=\tilde{\rho}_{ee}^{\left( 1\right)}~$ and $\rho _{ee}^{\left(
2\right)}(t)=\tilde{\rho}_{ee}^{\left( 2\right)}$.\newline
Inserting the solutions in above set of Eqs. (18a-18g) we have the following
closed set of equations for the steady state dynamics

\begin{subequations}
\begin{equation}
\left[ i\left( \omega _{1}-\omega \right) +\gamma _{1}\right] \tilde{\alpha}%
_{1}+2i\chi ^{\left( 3\right) }\tilde{\alpha}_{1}^{\ast }\tilde{\alpha}_{2}%
\tilde{\alpha}_{3}=\epsilon _{p}  \label{19a}
\end{equation}
\begin{equation}
\left[ i\left( \omega _{2}-\omega ^{\prime }\right) +\gamma _{2}\right] 
\tilde{\alpha}_{2}+i\chi ^{\left( 3\right) }\tilde{\alpha}_{3}^{\ast }\tilde{%
\alpha}_{1}^{2}=\epsilon _{p}^{\prime}  \label{19b}
\end{equation}
\begin{equation}
\left[ i\left( \omega _{3}+\omega ^{\prime }-2\omega \right) +\gamma _{3}%
\right] \tilde{\alpha}_{3}+i\chi ^{\left( 3\right) }\tilde{\alpha}_{2}^{\ast
}\tilde{\alpha}_{1}^{2}=-if_{1}\tilde{\rho}_{ge}^{\left( 1\right) }--if_{2}%
\tilde{\rho}_{ge}^{\left( 2\right)}  \label{19c}
\end{equation}
\begin{equation}
\left[ i\left( \omega _{eg}^{\left( 1\right) }+\omega ^{\prime }-2\omega
\right) +\gamma _{eg}^{\left( 1\right) }\right] \tilde{\rho}_{ge}^{\left(
1\right) }=if_{1}^{\ast }\tilde{\alpha}_{3}y_{1}+ig^{\ast }y_{1}\tilde{\rho}%
_{ge}^{\left( 2\right)}  \label{19d}
\end{equation}
\begin{equation}
\left[ i\left( \omega _{eg}^{\left( 2\right) }+\omega ^{\prime }-2\omega
\right) +\gamma _{eg}^{\left( 2\right) }\right] \tilde{\rho}_{ge}^{\left(
2\right) }=if_{2}^{\ast }\tilde{\alpha}_{3}y_{2}+igy_{2}\tilde{\rho}%
_{ge}^{\left( 1\right)}  \label{19e}
\end{equation}
\begin{equation}
\gamma _{ee}^{\left( 1\right) }\tilde{\rho}_{ee}^{\left( 1\right) }=i\left(
f_{1}\tilde{\alpha}_{3}^{\ast }\tilde{\rho}_{ge}^{\left( 1\right)
}-f_{1}^{\ast }\tilde{\alpha}_{3}\tilde{\rho}_{eg}^{\left( 1\right) }\right)
+i\left( g\tilde{\rho}_{eg}^{\left( 2\right) }\tilde{\rho}_{ge}^{\left(
1\right) }-g^{\ast }\tilde{\rho}_{eg}^{\left( 1\right) }\tilde{\rho}
_{ge}^{\left( 2\right) }\right)  \label{19f}
\end{equation}
\begin{equation}
\gamma _{ee}^{\left(2\right)}\tilde{\rho}_{ee}^{\left( 2\right) }=i\left(
f_{2}\tilde{\alpha}_{3}^{\ast }\tilde{\rho}_{ge}^{\left( 2\right)
}-f_{2}^{\ast }\tilde{\alpha}_{3}\tilde{\rho}_{eg}^{\left( 2\right) }\right)
+i\left( g^{\ast }\tilde{\rho}_{eg}^{\left( 1\right) }\tilde{\rho}%
_{ge}^{\left( 2\right) }-g\tilde{\rho}_{eg}^{\left( 2\right) }\tilde{\rho}%
_{ge}^{\left( 1\right) }\right)  \label{19g}
\end{equation}
where $\tilde{\alpha}_{1},\tilde{\alpha}_{2},\tilde{\alpha}_{3},\tilde{\rho}
_{ge}^{\left( 1\right) },\tilde{\rho}_{ge}^{\left(2\right)},\tilde{\rho}
_{ee}^{\left( 1\right) }$ and $\tilde{\rho}_{ee}^{\left( 2\right) }$are
constants independent of time. $y_{i}=\left( \tilde{\rho}_{ee}^{\left(
i\right) }-\tilde{\rho}_{gg}^{\left( i\right) }\right) ~$are the population
inversion $\left( i=1,2\right) $ for both QDs.

Using Eqs.(19d) and (19e) in Eq.(19c), we obtain the steady state value for $%
\tilde{\alpha}_{3}$ as follows. 
\end{subequations}
\begin{equation}
\tilde{\alpha}_{3}=\frac{i\chi ^{\left( 3\right) }\left( \beta _{1}\beta
_{2}+y_{1}y_{2}\left\vert g\right\vert ^{2}\right) }{\left( y_{1}\left\vert
f_{1}\right\vert ^{2}\beta _{2}+y_{2}\left\vert f_{2}\right\vert ^{2}\beta
_{1}\right) +iy_{1}y_{2}\left( f_{1}f_{2}^{\ast }g^{\ast }+f_{1}^{\ast
}f_{2}g\right) -\varepsilon _{3}\left( \beta _{1}\beta
_{2}+y_{1}y_{2}\left\vert g\right\vert ^{2}\right) }\tilde{\alpha}_{2}^{\ast
}\tilde{\alpha}_{1}^{2}
\end{equation}

where the short hand notations are $\varepsilon _{1}=\left[ i\left( \omega
_{1}-\omega \right) +\gamma _{1}\right] ,~\varepsilon _{2}=\left[ i\left(
\omega _{2}-\omega ^{\prime }\right) +\gamma _{2}\right] ,~\varepsilon _{3}=%
\left[ i\left( \omega _{3}+\omega ^{\prime }-2\omega \right) +\gamma _{3}%
\right] ~$\ and $\beta _{1}=\left[ i\left(\omega _{eg}^{\left( 1\right)
}+\omega ^{\prime }-2\omega \right) +\gamma _{eg}^{\left( 1\right) }\right]
~ $\ and $\beta _{2}=\left[ i\left( \omega _{eg}^{\left( 2\right) }+\omega
^{\prime }-2\omega \right) +\gamma _{eg}^{\left( 2\right) }\right] .$

\subsection{Super enhancement of FWM process}

\subsubsection{Single QD case}

In case of a single QD coupled to the gold grating, $f_{2}=g=g^{\ast}=0$ and 
$f_{1}=f$, we get the steady state value of $\tilde{\alpha}_{3}\ $ from
Eq.(20)as

\begin{equation}
\tilde{\alpha}_{3}=\frac{i\chi ^{\left( 3\right)}\tilde{\alpha}_{2}^{\ast} 
\tilde{\alpha}_{1}^{2}}{\frac{\left\vert f\right\vert ^{2}y}{\left[i\left(
\omega _{eg}+\omega ^{\prime }-2\omega \right) +\gamma _{eg}\right]}-\left[
i\left( \omega _{3}+\omega ^{\prime }-2\omega \right) +\gamma _{3}\right]}
\end{equation}
which coincides exactly with Eq.(9) where the imaginary part of the first
term $\frac{\left\vert f\right\vert ^{2}y}{\left[ i\left(\omega _{eg}+\omega
^{\prime }-2\omega \right) +\gamma _{eg}\right] }$ can be arranged to cancel
the $i\left( \omega _{3}+\omega ^{\prime }-2\omega \right)$ factor in the
second term of the denominator and this gives enhancement of FWM as also
discussed in previous section also.

\subsubsection{Coupled QDs case}

As compared to single QD case, the denominator of Eq.(20) can (in principal)
be arranged down to very low values in order to enhance $\tilde{\alpha}_{3}$
to much higher values. In this case, denominator has 3 complex $%
\left(f_{1},f_{2},g\right)$ and 2 real $\left(\omega _{eg}^{\left(1\right)}
,\omega _{eg}^{\left(2\right)}\right)$ parameters which can be tuned
independently.

We obtain  nearly $1200$ times enhancement by comparing the steady state
values of $\left\vert\tilde{\alpha}_{3}\right\vert ^{2}$ that is the
intensity of FWM process calculated from time evolution of Eqs.$\left(
18a-18g\right)$ for the chosen set of parameters as shown in Fig.$\left(
4\right)$. Here, frequency of second QD is kept constant and first one is varying. For the decay rates of grating modes  in between $0.01$, we get enhancement in FWM Intensity around 1200-1600 times as compared to case of single QD discussed in previous section.
\begin{figure}[hbt!]
\centering
{\includegraphics[width=0.5\textwidth]{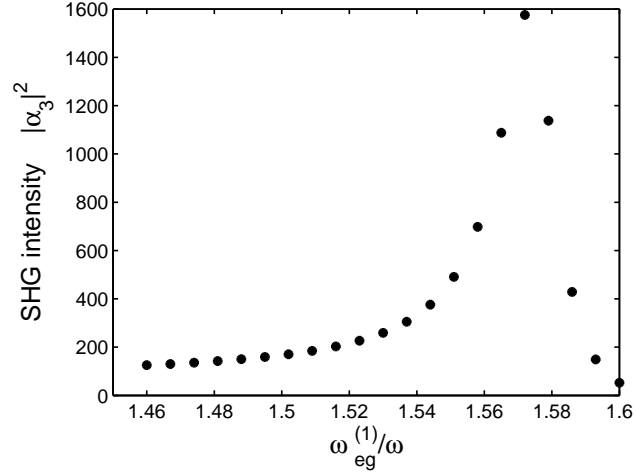}}
\caption{The enhancement of the FWM process in case of coupled QDs. The FWM
mode $\hat{a}_{3}$ is far-off resonant to the FWM condition $\left(
\omega _{3}=1.90\omega \right) $. The FWM process can be carried
closer to resonance by arranging the quantum level spacing of first QD to $
\left(\omega _{eg}^{\left(1\right)}\approx 1.5732\omega
\right)$, while second QD $\left(\omega _{eg}^{\left(2\right)
}= 1.5810\omega\right)$ being fixed . The conversion is enhanced nearly by $1600$ times. Decay rates of the grating modes are taken as $\gamma _{1}=\gamma _{2}=\gamma_{3}=0.01\omega $. We use $\chi ^{\left( 2\right)}=0.00001\omega$ and $f_{1}=f_{2}=0.1909\omega$, $g=\left(0.1000 + 0.0101i\right)\omega$,$
\gamma _{ee}^{\left(1\right)}=\gamma _{ee}^{\left(
2\right)}=0.00001\omega$, $\gamma _{eg}^{\left(1\right)}=\gamma
_{ee}^{\left(1\right)}/2$, $\gamma _{eg}^{\left( 2\right)}=
\gamma _{ee}^{\left(2\right)}/2$ for our numerical simulations.}
\label{Fig 4}
\end{figure}
\section{conclusion}
It is well demonstrated that the presence of a quantum
emitter with a smaller decay rate changes the optical
response of coupled grating dramatically. Due to the destructive
interference of the (hybridized) absorption paths, Four wave mixing(FWM) process 
can be suppressed at the resonance frequency of the quantum emitter.
We demonstrate that a similar path interference effect
can be adopted to both suppress and enhance the nonlinear
Four wave mixing processes (FWM) in a grating surface. A quantum
emitter is coupled with the electromagnetic modes of a gold grating. 
 We found that the FWM process can be suppressed over 10 orders
of magnitude. Such an suppression can be achieved by carefully choosing the coupling strengths and the energy level spacing for quantum emitters. When $\left(\omega _{eg}=2\omega -\omega
^{\prime}\right)$, the FWM intensity can be suppressed by several order of 
magnitude with respect to resonant value. On the other hand, the similar
interference effects can be also used to enhance the
nonlinear FWM intensity. The level spacing of the
single quantum emitter can be arranged so that the nonresonant terms get canceled.
In case of two coupled quantum emitters by arranging energy level spacing for quantum emitters in the same way like single quantum emitter, we have enhancement in FWM intensity upto the order of $10^{3}$.

\end{document}